\documentclass{article}

\PassOptionsToPackage{numbers, compress}{natbib}

\usepackage[preprint]{neurips_2024}

\usepackage{wrapfig}
\usepackage{epsfig}
\usepackage[utf8]{inputenc} 
\usepackage[T1]{fontenc}    
\usepackage{hyperref}       
\usepackage{url}            
\usepackage{booktabs}       
\usepackage{amsfonts}       
\usepackage{nicefrac}       
\usepackage{microtype}      
\usepackage[dvipsnames]{xcolor}         

\usepackage{amsmath}
\usepackage{diagbox}
\usepackage{graphics}
\usepackage{epsfig}
\usepackage{threeparttable}
\usepackage{xspace}
\usepackage{siunitx}
\usepackage{graphicx}
\usepackage{amssymb}
\usepackage{threeparttable}
\usepackage{wrapfig}
\usepackage{color}
\usepackage[normalem]{ulem}
\usepackage{multirow}
\usepackage{float}
\usepackage{amsfonts}
\usepackage{bm}
\usepackage{array}
\usepackage{colortbl}
\usepackage{tikz}
\usepackage{diagbox}
\usepackage{rotating}
\usepackage{booktabs}
\usepackage{overpic}
\usepackage{enumitem}
\usepackage{colortbl}
\usepackage[export]{adjustbox}
\usepackage{listings}
\usepackage{pifont}
\usepackage{makecell}
\usepackage{arydshln}

\newcommand{\ie}{\textit{i}.\textit{e}.}
\newcommand{\eg}{\textit{e}.\textit{g}.}

\usepackage{booktabs} 
\usepackage{algorithm}
\usepackage{amsmath}
\usepackage{amssymb}
\usepackage{mathtools}
\usepackage{amsthm}
\definecolor{mygray}{gray}{.92}

\DeclareMathOperator*{\argmin}{arg\,min}

\usepackage{makecell}
\usepackage{arydshln}
\usepackage[export]{adjustbox}
\usepackage{listings}
\usepackage{pifont}
\usepackage{diagbox}
\usepackage{graphicx}
\usepackage{amssymb}
\usepackage{threeparttable}
\usepackage{wrapfig}
\usepackage{xspace}
\usepackage{diagbox}
\usepackage{graphics}
\usepackage{epsfig}
\usepackage{threeparttable}
\usepackage{siunitx}
\usepackage{color}
\usepackage[normalem]{ulem}
\usepackage{multirow}
\usepackage{float}
\usepackage{amsfonts}
\usepackage{bm}
\usepackage{array}
\usepackage{rotating}
\usepackage{booktabs}
\usepackage{overpic}
\usepackage{enumitem}
\usepackage{colortbl}
\usepackage{multirow}
\usepackage{subcaption}
\definecolor{mygray}{gray}{.9}

\makeatletter
\newcommand{\thickhline}{%
	\noalign {\ifnum 0=`}\fi \hrule height 1pt
	\futurelet \reserved@a \@xhline
}
\makeatother

\usepackage[capitalize,noabbrev]{cleveref}
\usepackage{multirow}

\title{Diff-PIC: Revolutionizing Particle-In-Cell Nuclear Fusion Simulation with Diffusion Models}

\author{
  Chuan Liu \\
  University of Rochester\\
  \texttt{cliu81@ur.rochester.edu} \\
\And
  Chunshu Wu \\
  University of Rochester\\
  \texttt{cwu88@ur.rochester.edu} \\
\And
  Shihui Cao \\
  University of Rochester \\
  \texttt{scao5@ur.rochester.edu} \\
\And
  Mingkai Chen \\
  Rochester Institute of Technology\\
  \texttt{mxceec@rit.edu} \\
\And
  James Chenhao Liang \\
  Rochester Institute of Technology\\
  \texttt{jcl3689@rit.edu} \\
\And
  Ang Li \\
  Pacific Northwest National Laboratory\\
  \texttt{ang.li@pnnl.gov} \\
\And
  Michael Huang \\
  University of Rochester\\
  \texttt{michael.huang@rochester.edu} \\
\And
  Chuang Ren \\
  University of Rochester\\
  \texttt{chuang.ren@rochester.edu} \\
\And
  Dongfang Liu \\
  Rochester Institute of Technology\\
  \texttt{dongfang.liu@rit.edu} \\
\And
  Ying Nian Wu\\
  University of California, Los Angeles\\
  \texttt{ywu@stat.ucla.edu} \\
\And
  Tong Geng\thanks{Corresponding author} \\
  University of Rochester\\
  \texttt{tong.geng@rochester.edu} \\
}

\begin{document}

\maketitle

\begin{abstract}
    The rapid development of AI highlights the pressing need for sustainable energy, a critical global challenge for decades. Nuclear fusion, generally seen as an ultimate solution, has been the focus of intensive research for nearly a century, with investments reaching hundreds of billions of dollars. Recent advancements in Inertial Confinement Fusion have drawn significant attention to fusion research, in which Laser-Plasma Interaction (LPI) is critical for ensuring fusion stability and efficiency. However, the complexity of LPI upon fusion ignition makes analytical approaches impractical, leaving researchers depending on extremely computation-demanding Particle-in-Cell (PIC) simulations to generate data, presenting a significant bottleneck to advancing fusion research. In response, this work introduces Diff-PIC, a novel framework that leverages conditional diffusion models as a computationally efficient alternative to PIC simulations for generating high-fidelity scientific LPI data. In this work, physical patterns captured by PIC simulations are distilled into diffusion models associated with two tailored enhancements: (1) To effectively capture the complex relationships between physical parameters and corresponding outcomes, the parameters are encoded in a physically-informed manner. (2) To further enhance efficiency while maintaining high fidelity and physical validity, the rectified flow technique is employed to transform our model into a one-step conditional diffusion model. Experimental results show that Diff-PIC achieves 16,200$\times$ speedup compared to traditional PIC on a 100 picosecond simulation, with an average reduction in MAE / RMSE / FID of 59.21\% / 57.15\% / 39.46\% with respect to two other SOTA data generation approaches. 
\end{abstract}

\section{Introduction} \label{sec:intro}
Sustainable energy stands as one of the paramount challenges of our era, particularly with the rapid advancement of AI. The recent successful demonstration of fusion ignition~\citep{abu2024achievement} underscores the transformative potential of fusion as a sustainable energy source. In 2023 and 2024, the National Ignition Facility (NIF) achieved groundbreaking milestones, generating 3.4 MJ and 5.2 MJ of fusion energy, respectively, from 2.2 MJ input energy. Given that the estimated output could reach $\sim$120 MJ~\citep{suter2004prospects}, there is a growing demand for a deeper understanding of the fundamental science behind ignition efficiency, especially the physical mechanisms governing the interaction between the laser and the plasma emitted when the laser bombards the fuel pellet. However, the Laser-Plasma Interaction (LPI) is a complex multi-body problem, which traditionally relies on time-stepping method, particularly, Particle-in-Cell (PIC) simulations~\citep{tskhakaya2007particle,langdon2014evolution,sulsky1995application,arber2015contemporary,liewer1989general}. Despite being the preeminent standard for modeling the physics of LPI, PIC simulations are exceedingly intensive in computation, often requiring tens of millions of CPU hours, consuming millions of dollars in order to obtain meaningful outputs~\citep{germaschewski2016plasma,derouillat2018smilei,bastrakov2012particle}. The computational overhead of PIC simulations has become a daunting bottleneck in fusion research, raising the pressing need for innovative methodologies capable of generating high-quality scientific data with substantially reduced computational burden.

Over the years, numerous CPU-GPU implementations have been developed for PIC simulations~\citep{fonseca2002osiris,bowers2008ultrahigh,sgattoni2015optimising}. While invaluable, these efforts remain within the scope of the time-stepping approach that iteratively executes over infinitesimal time intervals, falling short in addressing the inherent computational overhead of long-term simulations. Recent advancements in generative AI -- diffusion models, however, present a novel approach to bypass the constraint. Diffusion models~\citep{sohl2015deep,ho2020denoising,song2019generative} have demonstrated exceptional capabilities in Computer Vision (CV)~\citep{song2023consistency, ramesh2022hierarchical,ho2022cascaded}, synthesizing highly complex data distributions that match real data with high fidelity. From the perspective of energy-based models~\citep{lecun2006tutorial,grathwohl2019your}, a diffusion model effectively constructs a highly complex energy field that governs the evolution of variables, analogous to the motion of particles in cells. In fact, diffusion models are rooted in the diffusion concept in physics, where particles move according to the energy field. This has sparked significant interest in their potential for generating scientific data, as recent applications of diffusion models in molecular dynamics simulations have demonstrated their promise in this domain~\citep{wu2023diffmd,petersen2023dynamicsdiffusion}.

Although diffusion models exhibit outstanding compatibility for generating PIC simulation data, two critical research gaps must be addressed. \ding{182} \textit{Physical soundness} must be ensured. 
In contrast to traditional PIC simulations that directly take continuous physical parameters as constraints, 
it remains unclear how diffusion models can effectively capture and distill complex physical patterns. \ding{183} \textit{Substantial efficiency improvement} must be achieved. The step-by-step denoising process in diffusion models is computationally demanding. Although the requirement for infinitesimal time intervals has been relaxed, the process still presents significant challenges similar to those faced by time-stepping methods, limiting the practicality of diffusion models as advanced alternatives for PIC simulations.


\begin{wrapfigure}{r}{0.45\textwidth}  
    \centering
    \vspace{-5.5mm}
    \includegraphics[width=0.45\textwidth]{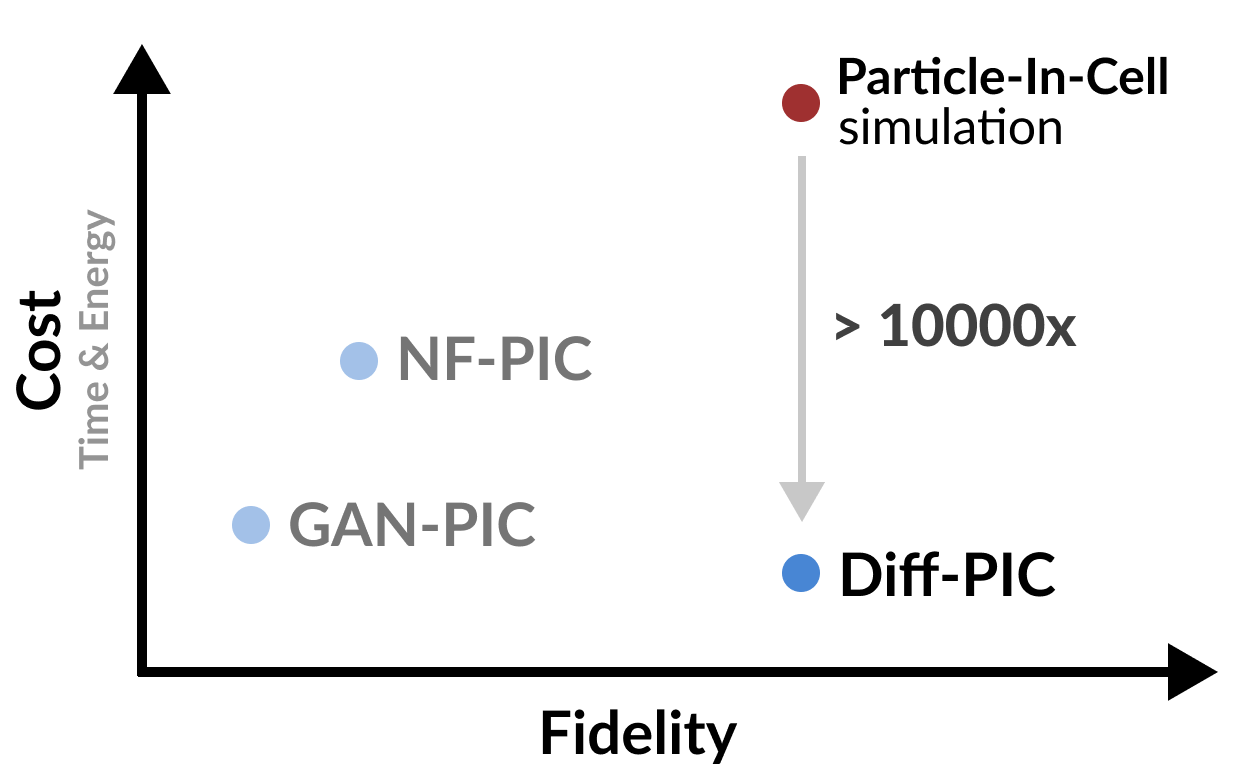}
    \vspace{-5.5mm}
    \caption{Overview of the proposed method.}
    \vspace{-2.5mm}
    \label{fig:intro}
\end{wrapfigure}
In light of these challenges, we propose a distillation framework for physical patterns, titled \textbf{Diff-PIC}, that leverages diffusion models to efficiently generate a snapshot of \textit{arbitrary} time, under \textit{arbitrary} simulation parameters within certain ranges. To specifically address the two challenges above: \ding{182} We develop a conditional diffusion model with a Physically-Informed Parameter Encoder. This encoder allows the model to capture the relationship between continuous physical parameters and PIC simulation data, distilling physical phenomena into Diff-PIC. \ding{183} We employ the rectified flow technique to eliminate the requirement for the multi-step denoising, further optimizing the runtime efficiency of Diff-PIC. As highlighted in Fig.~\ref{fig:intro}, orders-of-magnitude speedup is achieved compared to PIC simulations and other generative approaches including Generative Adversarial Networks (GAN)~\citep{aggarwal2021generative} and Normalizing Flow (NF)~\citep{zhang2021diffusion}, dubbed "GAN-PIC" and "NF-PIC".

In summary, our work represents the first known effort to tackle the imperative challenges associated with generating high-quality PIC simulation data for LPI using diffusion models renowned in the field of generative AI. The core contributions of our work include:
\begin{itemize}[leftmargin=*]
\setlength\itemsep{-0.5mm}
\item We propose Diff-PIC, a pioneering study that utilizes diffusion models as a computationally efficient alternative to PIC simulations (see \S\ref{subsec:overall}). By making all resources publicly available, we aim to establish Diff-PIC as a robust baseline and a valuable benchmark for the research community, thereby accelerating the advancement in scientific data generation for nuclear fusion research.
\item We develop a physically-informed conditional diffusion model (see \S\ref{subsec:PIPE}) that seamlessly integrates physical simulation parameters into the diffusion model. The designed condition encoder facilitates the generalization of effective simulation data within and beyond existing simulation parameters, endowing the model with robust generalization capabilities and adaptable transferability.
\item We implement the rectified flow technique to transform our model into a one-step diffusion model (see \S\ref{subsec:RFA}), thereby enhancing its efficiency in generating high-fidelity fusion data.
\item Experimental results in \S\ref{subsec:main_result} demonstrate that our method achieves a remarkable speedup of 16,200 times compared to traditional PIC simulations while preserving high fidelity and physical validity of the generated data -- 0.014, 0.019, 1.03 in terms of MAE, RMSE, and FID, respectively, with an average reduction of 59.21\% / 57.15\% / 39.46\% compared to two other SOTA generative models.
\end{itemize}

\section{Background}



\textbf{Inertial Confinement Fusion} (ICF) is a method of achieving controlled nuclear fusion by using intense energy pulses to compress and heat small fuel pellets~\citep{keefe1982inertial,betti2016inertial}, typically containing isotopes of hydrogen such as deuterium and tritium. This process unfolds the nuclear fusion reaction as delineated below:
\vspace{2mm}
\begin{equation}
{}^{2}\mathrm{H} + {}^{3}\mathrm{H} \rightarrow {}^{4}\mathrm{He} + {}^{1}\mathrm{neutron} + \textbf{Energy}.
\label{eq:fusion}
\end{equation}

Given the ubiquity of these hydrogen isotopes in the ocean, nuclear fusion holds immense potential to provide "near-infinite" energy by achieving the necessary temperature and pressure conditions to initiate fusion reactions, attaining a positive net energy gain (\ie, output energy surpassing the input). To optimize the sophisticated initiation of fusion, an advanced understanding of the underlying LPI mechanism is essential. For this purpose, PIC is considered a crucial tool to provide theoretical insights into LPI, due to its capability of predicting and interpreting physical phenomena. 


\textbf{Particle-in-Cell Simulations.} The PIC method is a computational technique widely used in the study of plasma physics and fusion energy research~\citep{tskhakaya2007particle, jiang2015affine,derouillat2018smilei}. Developed in the mid-20$th$ century, the PIC method has become a cornerstone in the simulation of complex plasma behaviors, enabling researchers to delve into the intricate dynamics of particles and electromagnetic fields~\citep{lange1978adpic,lewis1972comparison}. To highlight, PIC is especially useful in LPI studies~\citep{arber2015contemporary,strozzi2012fast,klimo2010particle}, which involves complex dynamics of electrons and ions. PIC simulations track the trajectories and interactions of these charged particles under the influence of electromagnetic fields, providing insights into both shock wave formation and heating mechanisms that are essential for ICF. 

In essence, PIC is an iterative time-stepping method applied to atomic particles such as electrons and ions. Within each iteration, particles are systematically arranged into discrete cells according to their spatial distribution, with their positions and velocities being updated over infinitesimally small time steps, typically on the scale of femtoseconds ($10^{-15}$ seconds). Unlike molecular dynamics widely applied in biology~\citep{geng2019applications, das2018molecular}, PIC simulations in LPI are characterized by intensive electromagnetic fields, which exert a significant influence on particle trajectories as follows:
\vspace{1mm}
\begin{equation}
    \frac{d\mathbf{v}_i}{dt} = \frac{q_i}{m_i} \left( \mathbf{E}(\mathbf{r}_i, t) + \mathbf{v}_i \times \mathbf{B}(\mathbf{r}_i, t) \right). \label{eq:PIC}
\end{equation}

For a particle at position $\mathbf{r}_i$ with charge $q_i$ and mass $m_i$, the equation describes how the velocity $\mathbf{v}_i$ evolves according to an energy field that consists of electric field $\mathbf{E}$ and magnetic field $\mathbf{B}$. To ensure accuracy, simulating LPI at the scale of mere hundreds of picoseconds (\ie, $10^{-10}$ seconds) requires hundreds of thousands of sophisticated PIC iterations. This imposes significant demands on computational storage and processing capabilities. As a result, the PIC methodology has emerged as a stringent bottleneck in fusion research, significantly constraining progress in this domain.


\textbf{Diffusion models} have emerged as a prominent class of generative models within the realm of artificial intelligence, offering an innovative methodology for the synthesis of high-fidelity data~\citep{ho2020denoising,ho2022cascaded,ramesh2022hierarchical}. Named after the physical concept, diffusion models use the idea of diffusion, which in physics refers to the random movement of particles from regions of high concentration to regions of lower concentration, often driven by thermal energy. 

In machine learning, diffusion models are generative models that progressively add noise to data in a forward process, then gradually remove it in the reverse process to generate new data. From the perspective of energy-based models, the reverse process can be seen as moving variables through an energy landscape, where the model transitions variables from high-energy, noisy states (where data are unstructured) to low-energy, clean states (where data are structured). In this sense, diffusion models generate data by allowing variables to evolve in a manner akin to particles diffusing within an energy field, granting the opportunity to effectively apply physical methods to evaluate the diffusion, specifically, Langevin dynamics:
\vspace{2mm}
\begin{equation}
    d\mathbf{x}=-\nabla_\mathbf{x} E(\mathbf{x})dt+\sqrt{2\beta}d\mathbf{W_t}
\end{equation}

The evolution of variables $\mathbf{x}$ is influenced by both the energy field $E(\mathbf{x})$ and the stochastic motion term $\mathbf{W_t}$ regulated by the diffusion coefficient $\beta$. Through the dynamics, variables move towards low energy regions, conceptually mirroring the behavior of particles in PIC simulations. This similarity encourages an exploration of diffusion models as potential alternatives to traditional PIC simulations.

\section{Diff-PIC} \label{sec:Diff-PIC}

\begin{figure}
    \vspace{-10pt}
    \centering
    \includegraphics[width=1.0\textwidth]{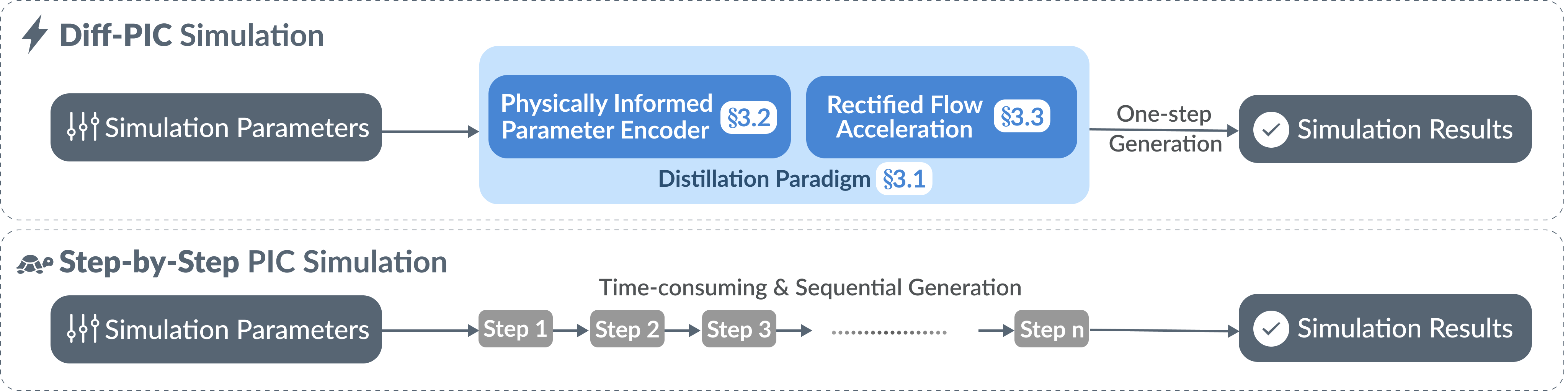}
    \caption{Workflow of the proposed Diff-PIC compared to traditional step-by-step PIC simulation.}
    \label{fig:overview}
    \vspace{-10pt}
\end{figure}

In this section, we introduce Diff-PIC, a physically-informed conditional diffusion model tailored for generating high-fidelity synthetic data for LPI in nuclear fusion. As illustrated in Fig.~\ref{fig:overview}, accepting physical parameters as inputs, the parameters are encoded and integrated with the model main body through the physically informed parameter encoder. Additionally, the rectified flow mechanism is adopted to break free from the time-stepping paradigm in both PIC simulation and traditional diffusion models, unlocking the model's capability to generate scientific data in one single step. 


\subsection{The Overall Distillation Framework} \label{subsec:overall}


Since LPI is governed by the behavior of electromagnetic field that dominates plasma dynamics, the tasks specifically assigned to Diff-PIC are to generate high-fidelity electric fields as representative cases under various physical parameters. Due to the polarization of the input laser, the resulting electric field only oscillates in a 2D plane, making 2D field descriptions sufficient to capture primary features of LPI. This approach is also a standard practice in PIC simulations. 

In this work, we focus on the following critical physical parameters in LPI: ``Electron Temperature ($T_e$) in $keV$,'' ``Ion Temperature ($T_i$) in $keV$,'' and ``Laser Intensity ($I$) in $W/m^2$.'' Additionally, to enable direct and versatile generation of the electric field at a specific simulation time (a snapshot), we include a user-defined parameter $t_{as}$ that represents the simulation time for an arbitrary snapshot. Through a standard learning process in diffusion models~\citep{ho2020denoising,song2021denoising}, the relationships between the four parameters and the resulting electric field are captured, as the model learns to progressively transform Gaussian noise into realistic electric fields. Once trained, the model takes the four parameters as inputs and produce electric field snapshots $\mathbf{E}(t_{as},\theta)$ corresponding to the specified parameters $\theta=\{T_e, T_i, I\}$. These generated snapshots can be used for multiple purposes, including data augmentation, parameter exploration, and studying the effects of different physical parameters. Furthermore, the diffusion model aids in the interpretation and visualization of complex plasma phenomena, providing valuable insights for researchers in the field of nuclear fusion. Essentially, the proposed techniques offer two advantages:
\begin{itemize}[leftmargin=*]
\setlength\itemsep{-0.5mm}
    \item \textit{Data Dependency Relaxation.} We treat snapshots from various parameter combinations as distinct distributions. This approach decouples data dependency in model training, enabling the model to efficiently learn from individual snapshots while generalizing across a wide array of scenarios.
    \item \textit{Systemic efficiency.} Unlike traditional PIC simulations that generate data sequentially over time, the proposed diffusion model can directly produce data for any target snapshot (see Fig.~\ref{fig:vis}). This non-sequential behavior allows for substantially more efficient data generation and analysis (see Table \ref{tab:speed}), enabling researchers to focus on specific times of interest without needing to simulate the entire LPI process from the beginning.
\end{itemize}

\subsection{Physically-Informed Parameter Encoder} \label{subsec:PIPE}

PIC simulations employ continuous physical parameters as inputs, which necessitate a seamless and continuous transition in the resulting synthesized data as input parameters are adjusted. Consequently, an encoder is considered exceptionally beneficial in this scenario, responsible for transforming domain-specific inputs into embeddings comprehensible by the model, meanwhile preserving the physical continuity of output snapshots.
In particular, these inputs comprise the simulation parameters $\theta$ and the target simulation snapshot $t_{as}$ that the conditional diffusion model aims to generate. 

Given the extensive range of physically feasible parameters and the limited data available during training, an optimal encoder must excel in both \textit{interpolation} and \textit{extrapolation} --- critical measures of the model's ability of generalization. Interpolation capability refers to the encoder's proficiency in generating suitable embeddings for new parameters that, although not encountered during training, lie between observed parameters. Extrapolation capability, conversely, pertains to generating embeddings for parameters that fall outside the range of those observed during training. Notably, both capabilities are indispensable for addressing the LPI problem, in order to cover a large enough, and fine-grained enough parameter space to provide sufficient insight into further LPI evaluations.

\begin{wrapfigure}{r}{0.4\textwidth}  
    \centering
    \vspace{-4.5mm}
    \includegraphics[width=0.4\textwidth]{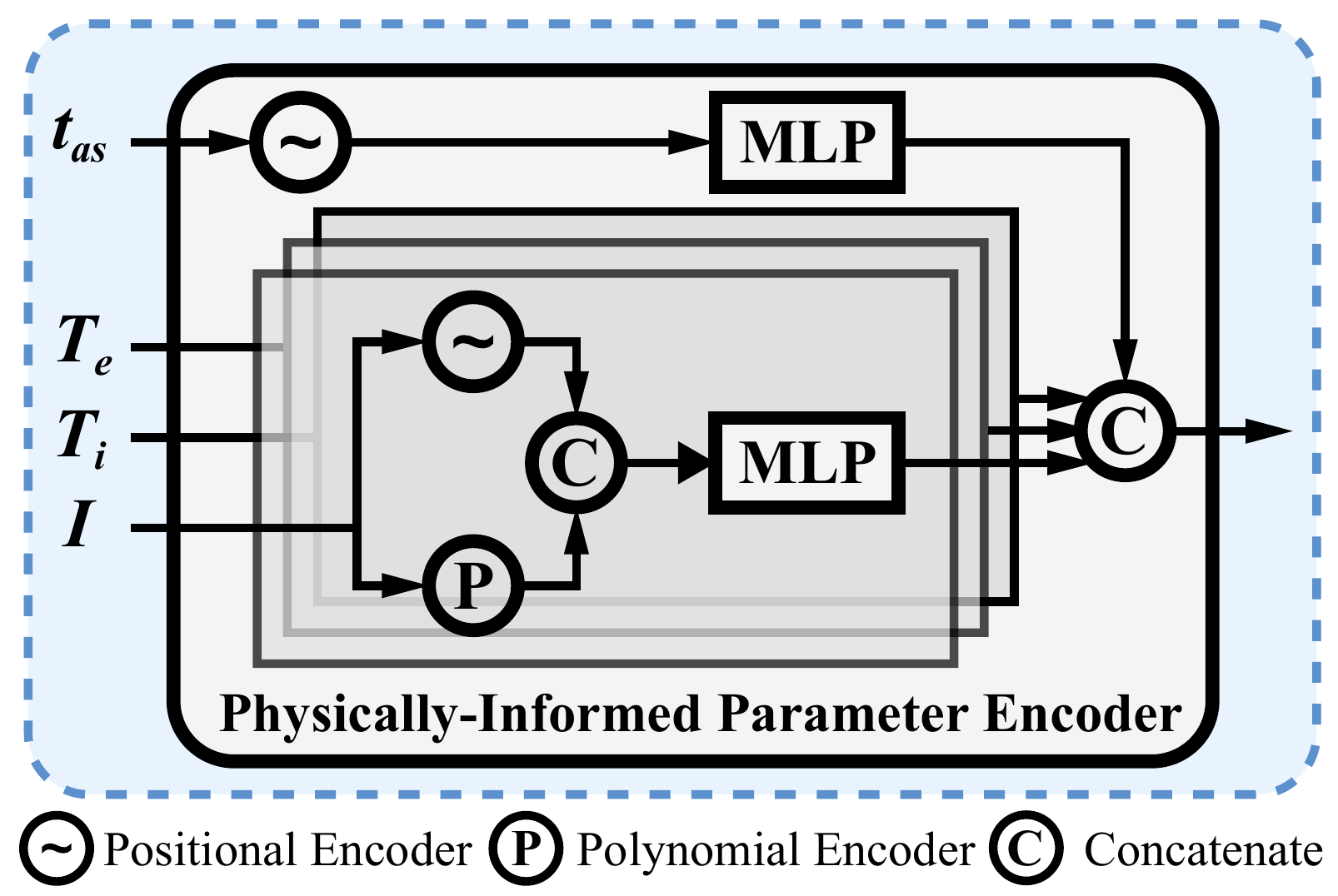}
    \vspace{-5.5mm}
    \caption{The proposed PIPE.}
    \vspace{-3.5mm}
    \label{fig:pipe}
\end{wrapfigure}
To meet the two essentials, we introduce a Physically-Informed Parameter Encoder (PIPE) as shown in Fig.~\ref{fig:pipe}. To encode the simulation parameters $\theta$, we incorporate two distinct types of encoders. For interpolation, we employ Positional Encoding~\citep{vaswani2017attention} (denoted ``$\sim$'' in a circle) to learn a continuous representation of input parameters, enabling smooth transitions between observed parameters. To augment extrapolation capability, we enhance the encoding process with a polynomial encoder (``P'' in a circle), utilizing transformation functions constructed as a linear combination of polynomial basis functions $f_i(\theta)$ of varying degrees: 
\vspace{2mm}
\begin{equation}
    \mathcal{P}(\theta) = \sum_{i=0}^{n} f_i(\theta),
\end{equation}

where $n$ denotes the maximum order of polynomial terms, and the polynomial $\mathcal{P}(\theta)$ can be chosen as Chebyshev polynomials and Legendre polynomials, based on the characteristics of the parameter space. This polynomial enhancement allows the encoder to generate plausible embeddings for parameters well beyond those encountered during training, ensuring robust performance across a broader spectrum of simulation scenarios.
Subsequently, we concatenate (``C'' in a circle) the embeddings from these two encoders and apply a Multi-Layer Perceptron (MLP) to further refine the embeddings. The MLP, with its trainable parameters, learns to combine and transform the concatenated embeddings, resulting in a more informative representation of the input parameters.
For encoding the simulation time step $t_{as}$, we utilize Positional Encoding ~\citep{vaswani2017attention} followed by an MLP layer. This approach is specifically chosen to learn continuous representations that facilitate smooth transitions between consecutive snapshots, thereby enhancing the model's temporal coherence.
In summary, this design offers the following advantages:

\begin{itemize}[leftmargin=*]
\setlength\itemsep{-0.5mm}
    \vspace{-2.0mm}
    \item \textit{Algorithmic generalization.} PIPE improves the generalizability of the conditional diffusion model (see Table~\ref{tab:extrapolation_e1}). The dual-encoding strategy captures non-linear relationships by incorporating both positional and polynomial encodings, empowering the model to adeptly manage a diverse array of simulation parameters, ranging from scenarios encountered during the training phase to parameters that lie beyond the spectrum of the training data.
    \item \textit{Adaptive transferability.} By fine-tuning the pre-trained model with a new dataset, this methodology facilitates adaptation to other fields where precise and efficient simulations are imperative for deciphering intricate physical phenomena.
\end{itemize}

\vspace{-3mm}

\subsection{Rectified Flow-Based Acceleration} \label{subsec:RFA}

To enable rapid generation of high-fidelity synthetic data, we employ the Rectified Flow Acceleration (RFA) technique in model optimization. Based on the principles of rectified flow~\citep{liu2022flow,esser2024scaling,liu2024instaflow}, RFA converts the original complex denoising trajectory from initial noise $\epsilon$ to the target electric field snapshot $\mathbf{E}(t_{as})$ into a direct and straight path -- the shortest route between two distributions.
During training, RFA minimizes the following objective function with straight ordinary differential equations:
\begin{equation}
\label{eq:rfa}
    \argmin_{\zeta} \quad \mathbb{E}\left [\int_{0}^{1}\left \| (\mathbf{E}(t_{as},\theta) - \epsilon) - \zeta(\mathbf{E}_t, t ~|~ t_{as}, \theta) \right \| ^2 dt \right  ],
\end{equation}
where $\mathbf{E}_t = t\mathbf{E}(t_{as},\theta)+(1-t)\epsilon$ denotes the linear interpolation between $\mathbf{E}(t_{as},\theta)$ and $\epsilon$ across the diffusion timeline, with $t$ ranging from 0 to 1. The score-based model $\zeta$, approximated using the U-Net, defines the learned trajectory. By minimizing the expectation of the squared deviations between the straight path $\mathbf{E}(t_{as},\theta) - \epsilon$ and the learned trajectory $\zeta(\mathbf{E}_t, t ~|~ t_{as}, \theta)$, RFA promotes the adoption of the shortest and most direct path in the denoising process, thus significantly reducing the denoising time.
Once this time-dependent score-based model $\zeta$ is trained, we further straighten the learned trajectories through an interactive reflow procedure ~\citep{liu2024instaflow}.

In summary, the RFA module provides additional benefits for our paradigm:
\begin{itemize}[leftmargin=*]
\setlength\itemsep{-0.5mm}
    \vspace{-2.0mm}
    \item \textit{Streamlined Denoising Process.} RFA significantly accelerates the denoising process (see Table~\ref{tab:speed}) by converting the complex data trajectory from initial noise to the target snapshot into a direct denoising step. By distilling the typically winding diffusion path into the shortest route between two distributions, RFA greatly reduces the time required for generating high-fidelity synthetic data.
    \item \textit{Robust Optimization.}
    RFA leverages the principles of rectified flow to minimize deviations between the winding data trajectory and the optimal, shortest path. This direct path approach reduces the possibility of error accumulation that can occur with more winding, iterative methods.
\end{itemize}
\vspace{-3mm}

\section{Evaluation}

\subsection{Experimental Setup} \label{subsec:setup}

\textbf{Datasets.}
We provide a new dataset comprising 6,615 simulations across varied physical simulation parameters, each containing 80 snapshots of electric fields along two orthogonal directions denoted E1 and E2. The data were generated by OSIRIS~\citep{fonseca2002osiris}, a well-established PIC simulation software suite. The dataset covers diverse parameters, including $T_e$, $T_i$, and $I$, all of which are critical parameters influencing the resultant electric fields. To foster further advancements in fusion and scientific data generation research, we will release the dataset publicly. 

\textbf{Metrics.} To validate physical soundness, Mean Absolute Error (MAE) and Root Mean Squared Error (RMSE) are used to evaluate the electric field difference and the energy difference between the Diff-PIC generated and the ground truth produced by PIC simulations. To better demonstrate the relative error, the dataset is normalized to [0,1]. To further evaluate the difference in the generated and the ground truth data distributions, the Fréchet Inception Distance (FID) metric is also employed, reflecting the fidelity of the electric fields produced by DiffPIC. 


\textbf{Baselines.} We compare Diff-PIC with two other SOTA generative models, Generative Adversarial Networks~\citep{karras2020analyzing} and Normalizing Flow~\citep{zhang2021diffusion}. The baseline models are implemented based on the setups provided in the original papers. Since neither of them originally supports learning meaningful embeddings for the physical parameters, for fair comparison, we equip them with the proposed PIPE to establish two baselines: GAN-PIC and NF-PIC. 

\textbf{Diff-PIC configurations.} The architectural foundation of our model is predicated on the U-Net framework~\citep{ronneberger2015u}, incorporating a series of three down-sampling blocks followed by three up-sampling blocks. The training regimen of our model encompasses an extensive 600 epochs, employing a batch size of 64, a configuration empirically validated to secure model convergence. Furthermore, the training protocol adheres to a fixed learning rate of $5\times10^{-4}$, optimized via the Adam optimizer~\citep{kingma2014adam}. The optimization objective is delineated in Eq.~\ref{eq:rfa}. In addition, in terms of the encoders, the positional encoding mechanism yields a 16-dimensional embedding from a single input, while the polynomial encoder operates with a highest polynomial order of 4. 

\begin{figure}[t]
    \centering
    \includegraphics[width=\linewidth]{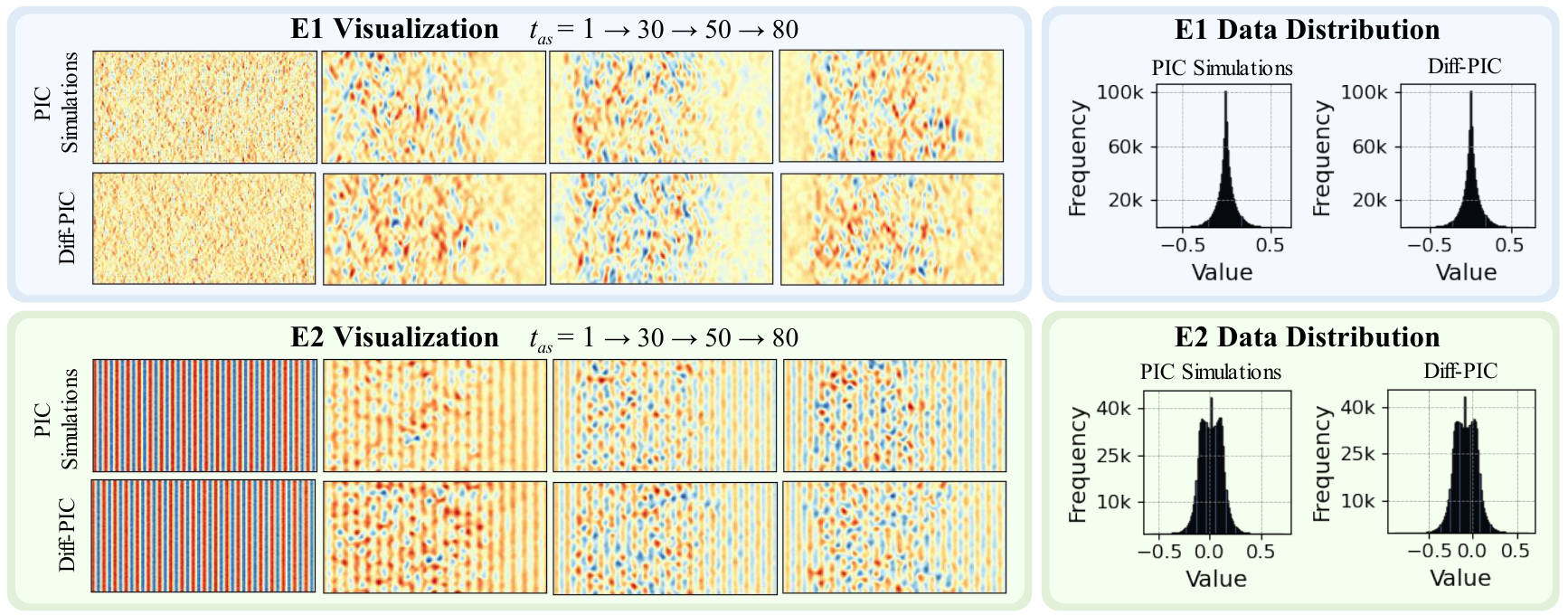}
    \vspace{-4mm}
\caption{\textbf{Visualization and Comparison} of PIC simulations and Diff-PIC.}
    \label{fig:vis}
    \vspace{-2mm}
\end{figure}

\subsection{Main Results} \label{subsec:main_result}

In this section, we evaluate Diff-PIC on three key aspects for comprehensive comparisons. (1) The interpolation ability and extrapolation ability. (2) The physical validity of the Diff-PIC generated data. (3) The speedup and power efficiency compared to traditional PIC simulations. 

\begin{table*}
\setlength{\tabcolsep}{3pt}
\footnotesize
\centering
\caption{\textbf{Quantitative results} for interpolation evaluation.}
\vspace{-1mm}
\resizebox{1\textwidth}{!}{
\begin{tabular}{l || c c c | c c c | c c c | c c c} 
       \hline
								\thickhline
        \vspace{-1.mm}
\multirow{2}{*}{\hspace{-1mm}\vspace{-1.5mm}\textbf{Method}} & \multicolumn{3}{c}{\textbf{Training Set for E1}} & \multicolumn{3}{c}{\textbf{Testing Set for E1}} & \multicolumn{3}{c}{\textbf{Training Set for E2}} & \multicolumn{3}{c}{\textbf{Testing Set for E2}} \\ 
\cmidrule(lr){2-4} 
\cmidrule(lr){5-7}
\cmidrule(lr){8-10}
\cmidrule(lr){11-13}
&  MAE$\downarrow$ & RMSE$\downarrow$ & FID$\downarrow$  & MAE$\downarrow$ & RMSE$\downarrow$ & FID$\downarrow$  & MAE$\downarrow$ & RMSE$\downarrow$ & FID$\downarrow$  & MAE$\downarrow$ & RMSE$\downarrow$ & FID$\downarrow$\\
\hline \hline
\textbf{GAN-PIC} & 4.59e-2 & 5.31e-2 & 2.32& 4.73e-2 & 5.84e-2 & 2.51 & 1.82e-2 & 2.07e-2 & 0.973 & 1.97e-2 & 2.18e-2 & 1.03\\ 
\textbf{NF-PIC} & 4.46e-2 & 5.12e-2 & 2.06 & 4.61e-2 & 5.35e-2 & 2.42 & 1.70e-2 & 2.03e-2 & 0.914 & 1.86e-2 & 2.45e-2 & 0.986\\ 
\textbf{Diff-PIC}  & 1.56e-2 & 2.67e-2 & 1.21 & 1.68e-2 & 2.29e-2 & 1.62 & 7.95e-3 & 9.32e-3 & 0.328 & 8.26e-3 & 1.03e-2 & 0.341 \\ 
\hline 
\end{tabular}
\label{tab:interpolation}
}
\vspace{-2mm}
\end{table*}

\textbf{Interpolation and Extrapolation. }To evaluate the interpolation capability of Diff-PIC, we sample a specified range for each simulation parameters ($T_e$, $T_i$, and $I$), totally 500 simulations and $500\times80=40,000$ snapshots. Then, we randomly split these 500 simulations into training and testing set with the ratio of 80\% and 20\%. We train Diff-PIC and baselines on the training set, and report the performance of Diff-PIC on the training and testing set in Table~\ref{tab:interpolation}. The reasonably low MAE, RMSE, and FID scores indicate that the proposed Diff-PIC is able to synthesize high-quality scientific data similar to what PIC generates, meanwhile significantly outperforming baselines on all three metrics. On average, Diff-PIC achieves 59.25\% reduced MAE, 57.77\% reduced RMSE, and 49.21\% reduced FID for both testing sets, compared to the baselines.

In addition, Fig.~\ref{fig:vis} compares the results for one randomly selected simulation produced by Diff-PIC and PIC respectively in the testing set. Throughout the snapshots ($t_{as}=1\rightarrow30\rightarrow50\rightarrow80$ for example), the distributions of the synthetic data closely follow the ground truths, indicating that physical continuity is maintained over time. Additionaly, the distributions on the right demonstrate that our proposed model successfully captures the data distributions in the ground truths. 
 
In terms of extrapolation capability evaluation, we progressively extend the range of physical parameters from the range selected for training. In particular, the ranges are extended by 10\% and 20\% respectively as case studies. The results on Table~\ref{tab:extrapolation_e1} demonstrate that Diff-PIC achieves approximately only 2\% relative absolute error in extrapolation tasks, significantly outperforming the other generative counterparts. On average, Diff-PIC achieves 59.16\% reduced MAE, 56.53\% reduced RMSE, and 29.70\% reduced FID for all test cases, compared to the baselines.

\begin{table*}[t]
\setlength{\tabcolsep}{3pt}
\footnotesize
\centering
\caption{\textbf{Quantitative results} for extrapolation evaluation.}
\vspace{-1mm}
\resizebox{1\textwidth}{!}{
\begin{tabular}{l || c c c | c c c | c c c | c c c} 
       \hline
								\thickhline
        \vspace{-1.5mm}
\multirow{2}{*}{\hspace{-1mm}\vspace{-1.5mm}\textbf{Method}} & & \multicolumn{1}{c}{\textbf{E1 10\%}} &&& \multicolumn{1}{c}{\textbf{E1 20\%}} &&& \multicolumn{1}{c}{\textbf{E2 10\%}} &&& \multicolumn{1}{c}{\textbf{E2 20\%}} \\ 
\cmidrule(lr){2-4} 
\cmidrule(lr){5-7}
\cmidrule(lr){8-10}
\cmidrule(lr){11-13}
&  MAE$\downarrow$ & RMSE$\downarrow$ & FID$\downarrow$  & MAE$\downarrow$ & RMSE$\downarrow$ & FID$\downarrow$  & MAE$\downarrow$ & RMSE$\downarrow$ & FID$\downarrow$  & MAE$\downarrow$ & RMSE$\downarrow$ & FID$\downarrow$\\
\hline  \hline
\textbf{GAN-PIC} & 5.24e-2 & 6.32e-2 & 1.97 & 5.73e-2 & 6.98e-2&2.15 & 2.18e-2& 3.25e-2&1.15& 2.93e-2&3.89e-2 &1.29\\ 
\textbf{NF-PIC} & 4.74e-2 &5.31e-2 & 1.85 & 5.41e-2 & 6.46e-2 & 2.08& 1.90e-2 & 3.16e-2& 1.04 & 2.63e-2 & 3.42e-2 & 1.17\\ 
\textbf{Diff-PIC} & 1.83e-2 & 2.40e-2 & 1.74 & 2.18e-2 & 2.62e-2 &1.82 & 9.47e-3&1.36e-2 & 0.536& 1.13e-2& 1.85e-2& 0.673 \\ 
\hline 
\end{tabular}
\label{tab:extrapolation_e1}
}
\end{table*}

\begin{figure}[t]
    \centering
    \hspace{-1mm}\includegraphics[width=1\linewidth]{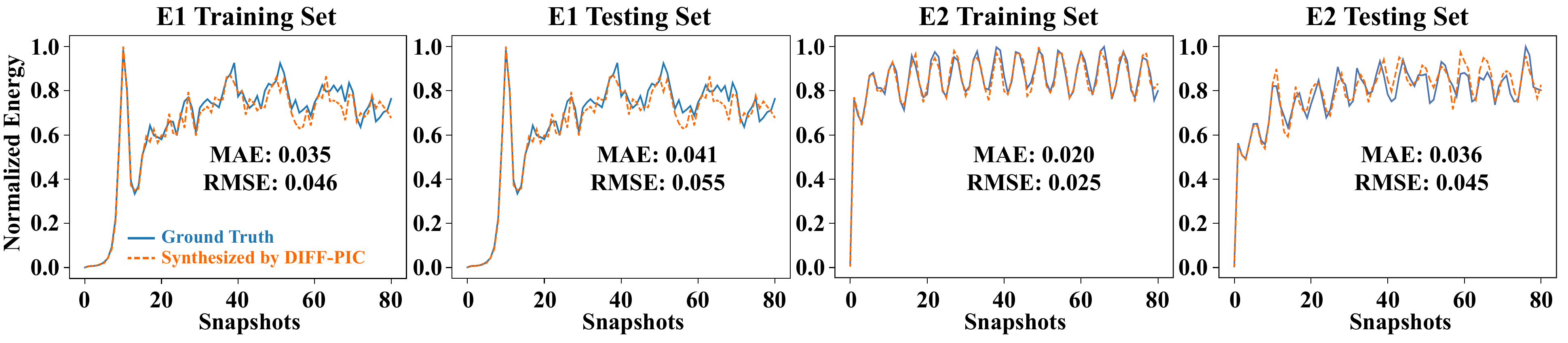}
    \caption{\textbf{Energy evaluation} of electric fields for training and test sets with 80 snapshots. }
    \label{fig:physic_inter}
    \vspace{-1mm}
\end{figure}

\textbf{Physical Validity.} In addition to the high quality of the synthetic data, we test the physical validity of the generated electric fields at each snapshot with energy, a fundamental concept used to describe the general property of a physical system. In particular, we randomly pick a simulation in testing set, the energy results of its electric fields are illustrated in Fig.~\ref{fig:physic_inter}. We observe that the pattern of the ground truth is captured by Diff-PIC, with fairly low MAE and RMSE incurred. Especially, for electric field E2, the oscillation in energy is properly preserved in the synthetic data. Furthermore, this evaluation highlights that Diff-PIC is capable of generating sequentially continuous data, demonstrating the effectiveness of the proposed distillation paradigm and the Physically-Informed Parameter Encoder.

\begin{table*}[t]
\setlength{\tabcolsep}{3pt}
\footnotesize
\centering
\caption{\textbf{Speedup} and \textbf{Energy Consumption Reduction}.}
\resizebox{1\textwidth}{!}{
\begin{tabular}{c || c | c c | c c | c c } 
       \hline
								\thickhline
								\rowcolor{mygray}
\textbf{Method} & PIC-100 ps & \textbf{Diff-PIC}-GPU & \textbf{Diff-PIC}-CPU & GAN-PIC-GPU & GAN-PIC-CPU & NF-PIC-GPU & NF-PIC-CPU\\
\hline \hline
\textbf{Speedup} & 1.00$\times$ & 1.62e4$\times$ & 519$\times$ & 1.56e4$\times$ & 523$\times$ & 9.21e2$\times$ & 24$\times$\\
\textbf{Energy Reduction} & 1.00$\times$ & 1.01e4$\times$ & 1.05e3$\times$ & 1.13e4$\times$ & 1.47e3$\times$ & 8.14e2$\times$ & 53$\times$\\
\hline
\end{tabular}
\label{tab:speed}
}
\vspace{-5mm}
\end{table*}

\textbf{Speedup and Power Efficiency.} In addition to the traditional PIC approach, baselines are also included in the comparison shown in Table~\ref{tab:speed}. The PIC simulations are run on the Perlmutter supercomputer in the National Energy Research Scientific Computing (NERSC) facility, with AMD EPYC 7763 CPUs. As essential fusion phenomena typically appear at approximately 100 ps, the costs of PIC simulation at 100 ps are selected as the reference across the comparisons. The GPU results for Diff-PIC and baselines are obtained on an Nvidia RTX 4090 GPU to demonstrate the availability of this approach to general users. The CPU results for these approaches are acquired on an Intel 13th Gen i9-13900KF CPU. The results highlight that Diff-PIC-GPU achieves over $10^4\times$ speedup versus traditional PIC simulation, as well as $10^4\times$ in terms of reduced energy cost.


\subsection{Discussion}

To provide further insights and highlight the value of this work, it is worth noting that our approach exhibits outstanding scalability compared to traditional PIC simulations. In PIC, for $N$ particles, the computing complexity can generally reach a formidable $TN\text{log}N$ with $T$ time steps. In contrast, Diff-PIC is not sensitive to the number of particles in space nor the number of time steps, since it focuses on generating the macroscopic data (\eg, electric field, which is usually considered more informative than individual particle status) directly for a specific time. For larger particle spaces, the speedup achieved by Diff-PIC can be easily improved by extra orders of magnitude, further accelerating the research of fusion, or other research areas involving large-scale PIC simulations.

\section{Related Work}
\textbf{Particle-in-Cell Simulations} have long been fundamental to modeling physical processes in fusion research~\citep{taccogna2017pic,garrigues2016negative}. However, the computational intensity of PIC simulations presents significant challenges~\citep{verboncoeur2005particle}. 
To mitigate these computational constraints, various methods have explored GPU and hybrid CPU-GPU acceleration technologies. Studies such as~\citep{abreu2010pic,burau2010picongpu,decyk2011adaptable,kong2011particle,suzuki2011acceleration} have utilized parallel computing, high memory bandwidth, and multiple processors to expedite simulations. Architecturally, the simulator optimized for the Kepler GPU architecture, as discussed in~\citep{shah2017novel}, underscores the potential of specific GPU architectures to enhance simulation efficiency. For more intricate simulations, research efforts like~\citep{xu2012discrete,chen2012efficient} have developed hybrid CPU-GPU implementations, \citep{wang2021hybrid} introduced a hybrid approach for multi-core and multi-GPU systems, highlighting the continuous integration and evolution of these technologies in advancing PIC simulations. Despite these advancements, these approaches remain reliant on the fundamental PIC framework, which may not completely address the computational burden due to the inherent algorithmic complexity of the PIC method.
In recent years, rapid advancements in deep learning have opened new pathways for accelerating scientific simulations. Machine learning-based approaches have emerged, such as predicting a vector space that approximates the PIC system solution~\citep{kube2021machine}, and learning the probability of interactions between potential collision pairs~\citep{bilbao2022towards}. However, the approach by~\citep{kube2021machine} depends on a pre-computed vector space and may not generalize well to novel scenarios, while~\citep{bilbao2022towards}'s method focuses on binary interactions, overlooking the complex many-body interactions in PIC simulations.

Contrastingly, our proposed method overcomes these limitations by employing a conditional diffusion model to distill the complex patterns captured by PIC simulations from a limited training dataset. Utilizing a time-dependent score-based model, our approach can efficiently generate high-fidelity synthetic data (see Fig.~\ref{fig:vis} and Table~\ref{tab:interpolation}) without the computational expense of traditional PIC algorithms (see Table~\ref{tab:speed}). This results in a significant reduction in computational cost while maintaining high simulation accuracy. Moreover, our method is highly adaptable, as it can be fine-tuned with minimal additional data to suit various physical parameters. This flexibility renders our approach suitable for a wide range of applications in fusion research and beyond, where efficient and accurate simulations are crucial for understanding complex physical phenomena.

\textbf{Diffusion Models in Scientific Research} have emerged as a formidable tool across a myriad of scientific domains. These models, which employ a stochastic process to incrementally convert a pristine data sample into a noise-distributed version and subsequently reverse this process. For instance, in materials science and chemistry, ~\citep{wu2023diffmd} introduced a diffusion model for molecular dynamics simulations, demonstrating the generalizability in generating molecular trajectories. ~\citep{arts2023one} presented an approach that integrates diffusion models with coarse-grained molecular dynamics to develop a new force field for simulating protein dynamics. By leveraging score-based generative models, they trained a model on coarse-grained structures to produce a force field that enhances the performance and realism of protein simulations without requiring force inputs during training. ~\citep{duan2023accurate} introduced an object-aware SE(3)-equivariant diffusion model for rapidly generating accurate 3D transition state structures, significantly reducing the computational burden typically associated with quantum chemistry calculations.
In astrophysics, diffusion models have been utilized to generate synthetic observations and simulate complex astrophysical phenomena. ~\citep{Smith_2022} proposed a diffusion model for generating realistic galaxy images, aiding in the analysis of large-scale sky surveys. Diffusion models have also found applications in climate science and Earth system modeling. For instance, ~\citep{oyama2023deep} employed a deep generative model to super-resolve spatially correlated multiregional climate data, enhancing the spatial resolution of global climate simulations, which is crucial for long-term climate projections and infrastructure development planning. ~\citep{li2024generative} explored the generative emulation of weather forecast ensembles with diffusion models, illustrating their effectiveness as scalable and cost-efficient alternatives to traditional ensemble forecasts, thus improving the reliability and accuracy of predictions for extreme weather events.
In particle physics, ~\citep{imani2024scorebased} introduced a diffusion model for generating high-quality Liquid Argon Time Projection Chamber (LArTPC) images, showcasing the model's ability to handle the challenges of sparse but locally dense particles.

While these studies highlight the expanding interest and application of diffusion models across various scientific domains, their potential within fusion research, specifically as an alternative to PIC simulations, remains underexplored. Our work aims to bridge this gap by proposing Diff-PIC, a conditional diffusion model that integrates the capability of PIC for generating high-fidelity synthetic data in fusion research. Diff-PIC leverages the inherent advantages of diffusion models to provide a computationally efficient alternative to traditional PIC simulations (see Table~\ref{tab:speed}).

\section{Conclusions}

This paper presents Diff-PIC, a pioneering approach that leverages the capabilities of diffusion models to generate high-fidelity synthetic data for LPI, offering a computationally efficient alternative to conventional PIC simulations in nuclear fusion research. By integrating a Physically-Informed Parameter Encoder and applying the Rectified Flow Acceleration, Diff-PIC significantly augments the diffusion model's capacity to manage diverse experimental parameters, thereby expediting the generation of high-fidelity synthetic data.
These advancements facilitate rapid, resource-efficient exploration of the design space, markedly diminishing the computational demands associated with PIC simulations. Our research not only catalyzes accelerated scientific discoveries within the realm of fusion research but also sets a novel precedent for the application of generative AI models in scientific simulations. Future investigations may focus on optimizing the distillation paradigm, harmonizing the simulation time $t_{as}$ with the diffusion time $t$, and refining the condition encoder to encompass a broader spectrum of physical parameters.

\bibliography{ref}
\bibliographystyle{splncs04}


\end{document}